\documentclass[preprint2]{aastex}

\def\sun{\hbox{$\odot$}}

\slugcomment{Accepted by the Astronomical Journal.}

\shorttitle{Optical-Infrared Light Curves of
AGB Stars}
\shortauthors{Smith et al.}

\begin{document}


\title{Phase Lags in the Optical-Infrared Light Curves
of AGB Stars}


\author{Beverly J. Smith}
\affil{Department of Physics, Astronomy, and Geology, East Tennessee
State University, Johnson City TN  37614}
\email{smithbj@etsu.edu}

\author{Stephan D. Price}
\affil{Air Force Research Laboratory, Space Vehicles Directorate, 
29 Randolph Road, Hanscom AFB, MA  01731}
\email{Steve.Price@hanscom.af.mil}

\author{Amanda J. Moffett}
\affil{Department of Physics, Astronomy, and Geology, East Tennessee
State University, Johnson City TN  37614}
\email{zajm17@etsu.edu}



\begin{abstract}
To search for phase lags in the optical-infrared light curves
of asymptotic giant branch stars, we have compared 
infrared data from 
the COBE DIRBE
satellite with optical light curves from the AAVSO and other
sources.
We found 17 examples of phase lags in the time of maximum in the
infrared vs.\ that in the optical, and 4 stars with no observed
lags.  There is
a clear difference between the Mira variables and the semi-regulars
in the sample, with the maximum in the optical preceding that in the
near-infrared in the Miras, while in most of the semi-regulars no
lags are observed.
Comparison to published theoretical models indicates that the phase
lags in the
Miras are due to strong titanium oxide absorption in the visual
at stellar maximum, and suggests that Miras pulsate in
the fundamental mode,
while 
at least some semi-regulars are first overtone 
pulsators.
There is a clear optical-near-infrared phase lag in the carbon-rich Mira V CrB;
this is likely due to C$_2$ and CN absorption variations in the
optical.

\end{abstract}





\section{Introduction}

Low mass stars ($<$8 M$_{\sun}$) are destined to pass through
the 
Asymptotic Giant Branch (AGB)
evolutionary stage.  AGB stars are
characterized by
physical pulsations and variations in the emitted light
\citep{wood97, olofsson99}, with high mass loss rates near
the end of the AGB stage
\citep{knapp85}.  
Dynamical models show that this mass loss
is driven by the pulsations, combined with non-LTE radiative 
relaxation and radiative acceleration of dust
\citep{bowen88, bowen91, willson00}.
Since these stars contribute significantly to
the enrichment of the interstellar medium \citep{busso99},
the study of AGB stars is vital to understanding the
lifecycle of the Galaxy as a whole.

To better understand the structure and evolution of AGB
stars, simultaneous optical and infrared monitoring can be useful.
Since 1933, it has been known that the maximum of
the light curves for Mira variables in the near-infrared
tends to lag that in the optical by about 10$\%$ of
the pulsation period
\citep{petit33, lockwood71, barnes73, maran77,
kerschbaum01, smith02, smith03, pardo04}.
These phase lags are still not well-quantified, and
only a few examples have been studied in detail.

With the advent of the Diffuse Infrared Background
Experiment (DIRBE) \citep{hauser98} instrument on
the Cosmic Background Explorer (COBE) satellite \citep{boggess92},
many high quality infrared light curves for AGB stars became
available \citep{smith02, smith03, knapp03}.
The DIRBE instrument provided photometry in 10 broadband
infrared filters (1.25, 2.2, 3.5, 4.9, 12, 25, 60, 100,
140, and 240 $\mu$m) with good temporal coverage
(100 $-$ 1000 data points in a 10 month period). 
This database was used to construct the COBE DIRBE
Point Source Catalog \citep{smith04}, in which DIRBE
photometry for essentially all of the infrared-bright
unconfused stars are tabulated and information on their 
DIRBE variability
is provided.
These light curves have been filtered to remove data points
affected by nearby companions, and stars with companions
in the DIRBE beam have been flagged and information about the 
companion is provided. 
In previous studies using DIRBE data, a few examples of
optical to near-infrared phase lags have been found
\citep{smith02, smith03, pardo04}.
In a few cases, 
offsets between the near-infrared and mid-infrared maxima were
also found \citep{smith02, smith03}.
In the current project, we expand these studies to a larger
sample.

\section{Method}

We pared down the 11,788 sources in the COBE DIRBE catalog
\citep{smith04}
to the 290 stars with DIRBE amplitudes
($\Delta$mag) greater than 5 times the uncertainty in 
the amplitude at
any of the five shortest
wavelength bands, that were unflagged
at that wavelength. We compared the light curves 
from these stars with the
visible curves from
the American Association
of Variable Star Observers 
(AAVSO\footnote{http://www.aavso.org}).
Of these stars,
199 were in the AAVSO database,
172 with AAVSO measurements
for the time
period in 1989 $-$ 1990 that DIRBE was operational
at cryogenic temperatures (JD = 2447874 $-$ 2448154) 
\citep{hauser98}.
High quality
visual light curves for RX Lep and UX Dra
were available
for this time
from
\citet{percy01} and \citet{buchler04}, respectively;
for these stars, these light curves were used instead of AAVSO data.
We also searched
the
Hipparcos database \citep{esa97} 
for optical data for the same
time period, 
but found no new examples of phase lags.

\section{Phase Lags}

The visible - infrared light curves for many of the 172 stars in
our sample indicate possible phase lags.  Unfortunately, 
the results for
most of these stars were ambiguous, either because 
their light curves
are incomplete 
or are confused with 
contamination from
nearby sources.  However, 17 of the stars have sufficient data to
clearly show optical-infrared phase effects while another 4 are
sufficiently well defined that we concluded that they exhibit 
no phase effects.  
In sources which were flagged at a wavelength in
the DIRBE Point Source Catalog, we 
carefully inspected the light curve and the DIRBE catalog
confusion notes on the companion, and only included
wavelengths where 
the effect of the companion was negligible.

The 21 sources in our final sample are listed in 
Table 1 in R.A. order, and their light curves are plotted
in Figures 1 $-$ 6.
Table 1 contains
the variability type and period from the General Catalog
of Variable Stars ((GCVS); \citet{GCVS}), 
and whether the star is carbon-rich, oxygen-rich, or with a carbon/oxygen
ratio C/0 $\approx$ 1 (type `S' stars).
Table 1 also includes periods determined from the optical data for
the 1200 day time period JD = 2447500 $-$ 2448700, which overlaps with
the DIRBE cryogenic period.

Note that the shape of the light curves in the optical
and the infrared often differ significantly, with the infrared
light curves tending to be more symmetrical than those in the optical.
As seen in Figure 1, many of the light curves are incomplete
and often a full pulsation period is not covered during the DIRBE
mission.  These properties limit the accuracy of the phase lag determinations.
We used two independent methods to determine the lags.
First, we determined the times of maxima by calculating
running weekly averages for the light curves. 
We estimated the uncertainty
on these maxima from the dates at which the brightness
in the weekly-averaged light curve has decreased from maximum by
1$\sigma$, where 1$\sigma$ is the rms in the original light curve
during the week of maximum.
Second, we estimated lags from cross-correlating the light curves
at the different wavelengths, after using a cubic
spline to re-grid the data to regular 1 day intervals.
In the cross-correlation, we only included data points brighter than
the mean brightness, since the shapes of the light curves are not
consistent from wavelength to wavelength.
Table 1 shows that the phase lags measured using these two methods
are consistent.  

There is a clear difference between the Miras and the semi-regulars
in Table 1.
All 16 of the Miras in Table 1 have lags in 
which the optical maximum precedes that in the
infrared, with
offsets of 20 $-$ $\ge$ 90 days.  
Of the 5 semi-regulars, four have no observed lags.
The fifth semi-regular, L$_2$ Pup (Figure 3),
has no lag or 
a possible
reverse lag (near-infrared leading optical) for one maximum,
and either a near-infrared/optical lag for the second maximum,
or an extra optical maximum not seen in the near-infrared.
In addition, the mid- and near-infrared light curves
of L$_2$ Pup show striking differences in
shapes.

The sole carbon-rich Mira in Table 1, 
V CrB, has a phase lag similar to those of the oxygen-rich Miras.
The two S type (C/O ratio $\sim$ 1) Miras R And and S Cas also have consistent
lags.
The sole carbon-rich semi-regular, UX Dra,
shows no lag.
There is also no clear distinction between the phase lags
of stars with silicate emission features and those with
featureless mid-infrared
spectra, as indicated by 
the 
infrared spectral types from the Infrared Astronomical Satellite
(IRAS) and the Infrared Space Observatory (ISO) 
\citep{kwok97, kraemer02}.
Note that because we are selecting sources based on the existence
of optical light curves, we are biased against very evolved, heavily obscured
AGB stars.    All of the oxygen-rich stars in Table 1 with available
infrared spectra
have silicate emission features 
or featureless mid-infrared
spectra; no stars with 
silicate absorption features are present in this sample.

Some examples of mid- to near-infrared phase shifts are
also present in Table 1 and Figure 1.  In particular, for R Hor,
S Pic,
and R Oct, the mid-infrared maxima precede those
in the near-infrared by $\sim$10 days.
There are no observed offsets in the minima of any star.

\section{Comparison to Theoretical Models }

To date, little discussion about the origin of phase lags 
has appeared in the literature.
\citet{alvarez98} suggested that optical-to-1-micron phase lags
are due to variations in the titanium oxide (TiO) and vanadium
oxide (VO) absorption in a stellar spectrum, with the TiO
strongly affecting broadband V measurements, and VO
contributing at $\approx$1 $\mu$m.
These molecules are
formed in different layers of the stellar atmosphere.  As the star
pulsates, shock waves travel through these layers,
creating and destroying these molecules at different times,
causing the opacities to vary with different phases.

To further investigate the origin of phase lags, and to understand
the observed differences between the lags in different types of stars,
in this section we compare our results to recent theoretical models of
AGB stars.

\subsection{Oxygen-Rich Stars}

A series of six time-resolved dynamical models 
for oxygen-rich stars
has been produced by
\citet{bessell96} and \citet{hofmann98}, supplemented by \citet{tej03}, 
\citet{ireland04a}, and \citet{ireland04b}.
These models 
include molecular opacities but not 
dust formation, and give
reasonably good fits to the 
optical and near-infrared (0.5 $-$ 2.5 $\mu$m) spectra of oxygen-rich
Miras \citep{tej03}.
The parameters of the six model stars in these papers 
are tabulated in \citet{tej03}.
They have periods from 320 $-$ 332 days, masses between 1 $-$ 2 M$_{\sun}$,
luminosities of 3470 $-$ 6310 L$_{\sun}$, and bolometric amplitudes
of 0.5 $-$ 1.2 magnitudes.
Four are fundamental mode pulsators, while two pulsate in the first
overtone.  

M. Scholz (2005, private communication) kindly provided us with
the theoretical spectra for these six models.  We convolved these spectra
with the DIRBE 1.25 $\mu$m, 2.2 $\mu$m, and 3.5 $\mu$m broadband
filter response functions, as well
as a standard V filter, to produce model broadband light curves.
In four of the six models, clear broadband optical-near-infrared phase lags
are seen, with the optical preceding the near-infrared by $\approx$0.1 $-$ 0.2
phase.  Inspection of the model spectra shows that this lag is caused
by strong TiO absorption in the optical near stellar maximum
(see Figure 8 in \citealp{bessell96}).
The infrared light curves tend to be more symmetric
than those in the optical, which often show a more gradual decline 
from maximum
(see, for example, S Pic in Figure 2).  
The infrared broadband fluxes
more closely trace the bolometric luminosity
of the star \citep{dyck74, blackwell90}, 
while the optical light curves are strongly attenuated 
by absorption.
The TiO absorption truncates the rise in the optical light at 
about phase $\approx$ 0, before the true maximum in the near-infrared.
Near-infrared VO and water absorption becomes stronger at later phases
than TiO (see Figure 8 in \citealp{bessell96}).
Water has strong features in the 2.5 $-$ 4 $\mu$m range \citep{aringer02},
which vary with pulsation cycle \citep{matsuura02}.

Interestingly, the four models with the phase lags are all
fundamental mode pulsators; the two first overtone models (models
E and O in \citealp{tej03}) do not
show phase lags.   
This supports the idea that Miras are fundamental mode
pulsators, while the semi-regulars without phase lags pulsate
in the first overtone.   Such a difference in mode
has been suggested before, based on luminosity-vs-period diagrams,
shock amplitudes,
and light curve shapes \citep{hill79, willson79, bessell96, willson00}.

The two models without phase lags have the lowest bolometric amplitudes,
0.5 and 0.7 magnitudes,
consistent with them being semi-regulars.
Miras have larger amplitudes and longer periods than
semi-regulars \citep{GCVS}.  Semi-regulars are better 
fit with hydrostatic models
than Miras \citep{loidl01, sudol02}, thus dynamical effects are more important
in Miras.

Since these models do not include a complete
treatment of dust, they only extend to 4 $\mu$m.  They therefore cannot be
used to investigate the observed mid-infrared/near-infrared phase shifts.
Such offsets may be due in part to dust emission in the
mid-infrared, powered by optical-UV heating.  This should be investigated
with 
more complete models including both molecular opacities as well
as dust formation, 
coupled to a dynamical model of stellar pulsations.
Including the silicate emission features would also be useful,
since these contribute to the 12 $\mu$m broadband flux 
and are known to vary with pulsation cycle in AGB stars
\citep{little96, creech97, monnier98, onaka02}.
As noted by \citet{bessell96} and \citet{ireland04b}, these models also 
do not treat deep
molecular absorption perfectly, thus the V and L (3.6 $\mu$m)
band fluxes are also somewhat uncertain.

\subsection{Carbon Stars}

Although TiO absorption can account for the observed phase lags
in oxygen-rich Miras, they cannot explain
the observed 
phase lag in the carbon Mira V CrB.
For this, carbon star models are needed.
The most complete models of carbon AGB stars to date
are those of S. H\"ofner and her collaborators
\citep{hofner98, hofner99, loidl99, hofner03, gautschy04}.
The latest models include time-dependent dynamics and
frequency-dependent radiative
transfer, as well as self-consistent time-dependent dust 
formation.
They provide good fits to carbon star spectra from 0.5 $-$ 5 $\mu$m, 
but show 
some discrepancies at longer wavelengths \citep{gautschy04}.
The model parameters vary between luminosities of 5200 $-$ 13,000 L$_{\sun}$,
stellar masses of 1 $-$ 2 M$_{\sun}$, effective temperatures of
2600 $-$ 3400K, periods between 148 and 525 days, amplitude velocities
of 2 $-$ 6 km~s$^{-1}$, and carbon-to-oxygen ratios C/O = 1.05 $-$ 2.0
\citep{hofner03, gautschy04}.
These models give mass loss rates of 0 $-$ 8 $\times$ 10$^{-6}$ M$_{\sun}$/yr,
and bolometric amplitudes of 0.10 $-$ 0.81 amplitudes
\citep{gautschy04}.

R. Gautschy-Loidl (2005, private communication) kindly provided
us with integrated broadband colors for these models.
About half of these models produce broadband optical-near-infrared
phase offsets, some in which the optical precedes the infrared,
and some the reverse.  The models with phase offsets tend to
have higher luminosities, cooler temperatures, higher mass loss
rates, higher bolometric amplitudes, and higher periods, but not in all
cases.  

The optical spectra of carbon stars are dominated by absorption
from C$_2$ and CN, while in the near-infrared 
C$_2$H$_2$, HCN, C$_3$ and CO are present
(see \citealp{loidl99}).  CN and C$_2$ form deeper in the
atmosphere than C$_2$H$_2$, C$_3$, and HCN \citep{loidl99}, thus
are affected at different times by a shock wave.
In the \citet{loidl99} model spectra, C$_2$H$_2$ absorption is minimum near
stellar maximum,
while 
the C$_2$ and CN absorption is weakest $\approx$0.1 phase before C$_2$H$_2$.
This is likely the cause of the observed phase lag in the Mira V CrB.
This should be confirmed with time-resolved spectroscopic
observations of V CrB.

According to \citet{bergeat05},
the two carbon stars in our sample, the Mira star
V CrB and the 
semi-regular UX Dra,
have bolometric luminosities of 6500L$_{\sun}$
and 10,600L$_{\sun}$, effective temperatures
of 2090K and 3090K, and mass loss rates of 
1.3 $\times$ 
10$^{-6}$ M$_{\sun}$/year
and 
3.7 $\times$ 
10$^{-7}$ M$_{\sun}$/year, respectively.
Their periods are 358 days and 168 days, respectively
\citep{GCVS}, and their DIRBE 2.2 $\mu$m amplitudes (which
approximate the bolometric amplitudes) are 0.8 and 0.1 magnitudes,
respectively.  Although there are no perfect matches to these 
particular sets
of parameters
in the published models,  except for the low temperature of 
V CrB each parameter is in the range covered by the models.
The lower temperature, higher mass loss rate, 
and 
larger amplitude of V CrB compared to the UX Dra is consistent with
it being more likely to have a phase lag, as observed.
New models with parameters better matched to these specific stars
would be helpful to compare with these light curves.

\section{Conclusions}

Using infrared light curves from DIRBE and optical
data from the AAVSO, we have found infrared-optical phase lags
in 17 stars, and no lags in 4 stars.
The Mira stars all show phase lags in which the optical maximum
precedes that in the near-infrared, while most of the semi-regulars
show no lags.
Comparison to published models shows that in the oxygen-rich Miras,
the phase lags are due to strong TiO absorption in the optical
near stellar maximum.  Published large amplitude,
fundamental mode models of oxygen-rich AGB
stars show phase lags,
while no lags are seen in models with small amplitude overtone mode pulsation. 
This is consistent with previous suggestions that Miras are
fundamental mode pulsators, and semi-regulars pulsate in the overtone 
mode.
The sole carbon Mira in the sample, V CrB,
shows an optical-first lag similar to that seen in the oxygen-rich
Miras; this is likely due to C$_2$ and CN absorption
in the visible.  





\acknowledgments

We would like to thank the COBE team for making 
this project possible.
We acknowledge with thanks the variable star observations from the 
AAVSO International Database contributed by observers worldwide 
and used in this research.
In particular,
we are grateful to Janet Mattei, the late Director
of the AAVSO, Elizabeth Waagen, the current Director,
and Mike Hauser, the Principal Investigator of
the DIRBE instrument on COBE.
We thank John Percy and Bob Cadmus for the use of their visual
light curves for RX Lep and UX Dra, and Michael Scholz
and Rita Gautschy-Loidl for providing us with their model results.
We are also grateful to 
J. M. Houchins for computer support, and
Mark Giroux, Lee Anne
Willson, Michael Scholz, Mike Ireland, Curt Struck, and the referee
for helpful comments.  The DIRBE Calibrated Individual Observations data product
was developed by the COBE Science Working Group and provided
by the National Space Science Data Center at NASA's Goddard
Space Flight Center.  This research has made use of the
SIMBAD Astronomical Database, operated at CDS, Strasborg, France,
as well as the NASA Astrophysics Data System at the Harvard-Smithsonian
Center for Astrophysics.
This research was funded by NASA LTSA grant NAG5-13079.

\clearpage


 \begin{figure}
\caption{
Filtered DIRBE light curves for the sample stars, compared to optical
light curves from the AAVSO.  The stars are in R.A. order, as in Table 1.
}
\end{figure}

\vfill
\eject

\clearpage

 \begin{figure}
\caption{
More filtered DIRBE light curves for the sample stars, compared to optical
light curves from the AAVSO.
}
\end{figure}

\vfill
\eject

\clearpage

 \begin{figure}
\caption{
More filtered DIRBE light curves for the sample stars, compared to optical
light curves from the AAVSO.  The RX Lep optical light curve is from 
\citet{percy01}.
}
\end{figure}

\vfill
\eject

\clearpage

 \begin{figure}
\caption{
More filtered DIRBE light curves for the sample stars, compared to optical
light curves from the AAVSO.
}
\end{figure}

\vfill
\eject

\clearpage

 \begin{figure}
\caption{
More filtered DIRBE light curves for the sample stars, compared to optical
light curves from the AAVSO.  The UX Dra optical light curve is from
\citet{buchler04}.
}
\end{figure}

\vfill
\eject

\clearpage

 \begin{figure}
\caption{
More filtered DIRBE light curves for the sample stars, compared to optical
light curves from the AAVSO.
}
\end{figure}


\clearpage








\begin{deluxetable}{|c|c|c|c|r|c|c|c|c|c|c}
\tabletypesize{\scriptsize}
\rotate
\tablecaption{DIRBE STARS WITH HIGH QUALITY OPTICAL AND INFRARED LIGHT CURVES\label{tbl-1}}
\tablewidth{0pt}
\tablehead{
\colhead{DIRBE NAME}&
\colhead{NAME}&
\colhead{Var}&
\colhead{C/O}&
\colhead{Period}&
\colhead{Period}&
\colhead{$\lambda$}&
\multicolumn{4}{c}{Optical/IR Lag}
\\
\colhead{}&
\colhead{}&
\colhead{Type}&
\colhead{Type}&
\colhead{(days)}&
\colhead{(days)}&
\colhead{($\mu$m)}&
\multicolumn{2}{c}{(weekly avg)}&
\multicolumn{2}{c}{(correlation)}
\\
\colhead{}&
\colhead{}&
\colhead{}&
\colhead{}&
\colhead{(GCVS)}&
\colhead{(this}&
\colhead{}&
\colhead{(days)}&
\colhead{(phase)}&
\colhead{(days)}&
\colhead{(phase)}
\\
\colhead{}&
\colhead{}&
\colhead{}&
\colhead{}&
\colhead{}&
\colhead{work)}&
\colhead{}&
\colhead{}&
\colhead{}&
\colhead{}&
\colhead{}
\\
}
\startdata
D00240197P3834373 & R And & M & S & 409 & 412 & 12 & $\ge$39 & $\ge$0.10 & $\ge$24 & $\ge$0.06 \\
D01194198P7236407 & S Cas & M & S & 612 & 598 & 3.5 & $\ge$23 & $\ge$0.04 & $\ge$9 & $\ge$0.02 \\
  &     &   &   &   &   & 4.9 & $\ge$25 & $\ge$0.04 & $\ge$9 & $\ge$0.02 \\
  &     &   &   &   &   & 12 & $\ge$28 & $\ge$0.05 & $\ge$30 & $\ge$0.05 \\
D02535274M4953225 & R Hor & M & O & 408 & 395 & 1.25 & 79$\pm$$^6_5$& 0.20$\pm$$^{0.02}_{0.01}$ & 97 & 0.25 \\
  &     &   &   &   &   & 2.2 & 103$\pm$$^7_5$& 0.26$\pm$$^{0.02}_{0.01}$ & 82 & 0.21 \\
  &     &   &   &   &   & 3.5 & 79$\pm$$^6_5$& 0.20$\pm$$^{0.02}_{0.01}$ & 77 & 0.19 \\
  &     &   &   &   &   & 4.9 & 79$\pm$$^6_5$& 0.20$\pm$$^{0.02}_{0.01}$ & 58 & 0.15 \\
  &     &   &   &   &   & 12 & 19$\pm$$^{32}_{19}$& 0.05$\pm$$^{0.08}_{0.05}$ & 38 & 0.10 \\
D03110304P1448000 & U Ari & M & O & 371 & 382 & 2.2 & $\ge$29 & $\ge$0.08 & $\ge$25 & $\ge$0.07 \\
D03524704M4549480 & U Hor & M & O & 348 & 353 & 1.25 & 103$\pm$$^{18}_{15}$& 0.29$\pm$$^{0.05}_{0.04}$ & 85 & 0.24 \\
  &     &   &   &   &   & 2.2 & 104$\pm$$^{19}_9$& 0.29$\pm$$^{0.05}_{0.03}$ & 90 & 0.25 \\
  &     &   &   &   &   & 3.5 & 104$\pm$$^{24}_{30}$& 0.29$\pm$$^{0.07}_{0.08}$ & 80 & 0.23 \\
  &     &   &   &   &   & 4.9 & 87$\pm$$^{18}_{10}$& 0.25$\pm$$^{0.05}_{0.03}$ & 57 & 0.16 \\
D05100884M6419044 & U Dor & M & O & 394 & 423 & 3.5 & 97$\pm$$^8_{41}$& 0.23$\pm$$^{0.02}_{0.10}$ & 60 & 0.14 \\
  &     &   &   &   &   & 4.9 & 97$\pm$$^5_{58}$& 0.23$\pm$$^{0.01}_{0.14}$ & 55 & 0.13 \\
  &     &   &   &   &   & 12 & 96$\pm$$^1_{39}$& 0.23$\pm$$^{0.01}_{0.09}$ & 50 & 0.12 \\
  &     &   &   &   &   & 25 & 45$\pm$$^7_{41}$& 0.11$\pm$$^{0.02}_{0.10}$ & 40 & 0.09 \\
D05105724M4830253 & S Pic & M & O & 428 & 422 & 1.25 & 80$\pm$$^7_{16}$& 0.19$\pm$$^{0.02}_{0.04}$ & 66 & 0.16 \\
  &     &   &   &   &   & 2.2 & 79$\pm$$^{19}_8$& 0.19$\pm$$^{0.05}_{0.02}$ & 88 & 0.21 \\
  &     &   &   &   &   & 3.5 & 78$\pm$$^8_{18}$& 0.18$\pm$$^{0.02}_{0.05}$ & 81 & 0.19 \\
  &     &   &   &   &   & 4.9 & 62$\pm$$^{18}_{20}$& 0.15$\pm$$^{0.05}_{0.05}$ & 51 & 0.12 \\
  &     &   &   &   &   & 12 & 9$\pm$$^{39}_{15}$& 0.02$\pm$$^{0.09}_{0.02}$ & 29 & 0.07 \\
  &     &   &   &   &   & 25 & 55$\pm$$^{62}_{78}$& 0.13$\pm$$^{0.15}_{0.18}$ & 28 & 0.07 \\
D05112286M1150566 & RX Lep$^a$ & SRb & 0 & 60 & 129 & 2.2 & 7$\pm$$^{10}_7$& 0.05$\pm$$^{0.08}_{0.05}$ & 4 & 0.03 \\
D05260609M8623179 & R Oct & M & O & 405 & 408 & 1.25 & 73$\pm$$^7_{14}$& 0.18$\pm$$^{0.02}_{0.03}$ & 52 & 0.13 \\
  &     &   &   &   &   & 2.2 & 69$\pm$$^{10}_2$& 0.17$\pm$$^{0.02}_{0.01}$ & 77 & 0.19 \\
  &     &   &   &   &   & 3.5 & 72$\pm$$^{11}_{28}$& 0.18$\pm$$^{0.03}_{0.07}$ & 55 & 0.13 \\
  &     &   &   &   &   & 4.9 & 59$\pm$$^{10}_{30}$& 0.15$\pm$$^{0.02}_{0.07}$ & 43 & 0.11 \\
D07133229M4438233 & L$_2$ Pup & SRb & 0 & 141 & 139 & 2.2 & $-$6$\pm$8/$-$13$\pm$$^{51}_{20}$& 0.04$\pm$0.06/$-$0.09$\pm$$^{0.37}_{0.22}$ & $-$10/30 & $-$0.07/0.22 \\
  &     &   &   &   &   & 3.5 & $-$6$\pm$$^{10}_7$/$-$9$\pm$$^{54}_{21}$& $-$0.04$\pm$$^{0.07}_{0.05}$/$-$0.06$\pm$$^{0.39}_{0.15}$ & $-$10/35 & $-$0.07/0.25 \\
  &     &   &   &   &   & 4.9 & $-$1$\pm$$^8_4$/8$\pm$$^{51}_{20}$& 0.01$\pm$$^{0.06}_{0.03}$/0.06$\pm$$^{0.37}_{0.14}$ & 9/$-$30 & 0.06/$-$0.22 \\
  &     &   &   &   &   & 12 & 0$\pm$$^9_5$/$-$53$\pm$$^{52}_{20}$& 0.00$\pm$$^{0.06}_{0.04}$/$-$0.38$\pm$$^{0.37}_{0.14}$ & $-$10/13 & $-$0.07/0.09 \\
D11491178M4145272 & X Cen & M & O & 315 & 312 & 2.2 & $\ge$63 & $\ge$0.20 & $\ge$51 & $\ge$0.16 \\
  &     &   &   &   &   & 3.5 & $\ge$71 & $\ge$0.23 & $\ge$46 & $\ge$0.15 \\
D12362346P5929128 & T UMa & M & O & 257 & 259 & 1.25 & 56$\pm$$^9_{19}$& 0.22$\pm$$^{0.03}_{0.07}$ & 52 & 0.20 \\
  &     &   &   &   &   & 2.2 & 44$\pm$$^8_{18}$& 0.17$\pm$$^{0.03}_{0.07}$ & 51 & 0.20 \\
  &     &   &   &   &   & 3.5 & 55$\pm$$^{14}_{21}$& 0.21$\pm$$^{0.05}_{0.08}$ & 50 & 0.19 \\
  &     &   &   &   &   & 4.9 & 40$\pm$$^9_{34}$& 0.15$\pm$$^{0.03}_{0.13}$ & 50 & 0.19 \\
D13294277M2316514 & R Hya & M & O & 389 & 435 & 1.25 & $\ge$40 & $\ge$0.09 & 36 & 0.08 \\
  &     &   &   &   &   & 2.2 & $\ge$43 & $\ge$0.10 & 55 & 0.13 \\
  &     &   &   &   &   & 3.5 & $\ge$46 & $\ge$0.11 & 58 & 0.13 \\
  &     &   &   &   &   & 4.9 & $\ge$33 & $\ge$0.08 & 53 & 0.12 \\
D14171992P6647391 & U UMi & M & O & 331 & 315 & 3.5 & 142$\pm$$^{23}_{56}$& 0.45$\pm$$^{0.07}_{0.08}$ & 99 & 0.31 \\
D14271640P0440414 & RS Vir & M & O & 354 & 360 & 1.25 & 57$\pm$$^{121}_6$& 0.16$\pm$$^{0.34}_{0.02}$ & 40 & 0.11 \\
  &     &   &   &   &   & 2.2 & 57$\pm$$^{120}_7$& 0.16$\pm$$^{0.33}_{0.04}$ & 44 & 0.12 \\
  &     &   &   &   &   & 3.5 & 57$\pm$$^{119}_{13}$& 0.16$\pm$$^{0.33}_{0.04}$ & 48 & 0.13 \\
  &     &   &   &   &   & 4.9 & 57$\pm$$^{120}_{14}$& 0.16$\pm$$^{0.33}_{0.04}$ & 41 & 0.11 \\
D15293454P7838003 & S UMi & M & O & 331 & 329 & 4.9 & 57$\pm$$^{12}_{34}$& 0.17$\pm$$^{0.04}_{0.10}$ & 43 & 0.13 \\
D15493131P3934178 & V CrB & M & C & 358 & 353 & 2.2 & 57$\pm$$^{29}_{47}$& 0.16$\pm$$^{0.08}_{0.13}$ & 48 & 0.14 \\
  &     &   &   &   &   & 3.5 & 58$\pm$$^{29}_{48}$& 0.16$\pm$$^{0.08}_{0.13}$ & 56 & 0.16 \\
  &     &   &   &   &   & 4.9 & 41$\pm$$^{38}_{48}$& 0.12$\pm$$^{0.11}_{0.14}$ & 49 & 0.14 \\
D16023917P4714250 & X Her & SRb & O & 95 & 106 & 2.2 & $-$7$\pm$$^7_{12}$/$-$28$\pm$$^{37}_{12}$& $-$0.07$\pm$$^{0.07}_{0.11}$/$-$0.26$\pm$$^{0.35}_{0.11}$ & $-$19 & $-$0.18 \\
  &     &   &   &   &   & 3.5 & 1$\pm$$^{29}_{22}$/$-$7$\pm$$^{43}_{38}$& 0.01$\pm$$^{0.27}_{0.21}$/$-$0.07$\pm$$^{0.41}_{0.36}$ & $-$17 & $-$0.16 \\
  &     &   &   &   &   & 4.9 & $-$1$\pm$$^7_{17}$/$-$2$\pm$39& $-$0.01$\pm$$^{0.07}_{0.16}$/$-$0.02$\pm$0.37 & $-$7 & $-$0.07 \\
D16481665P5748493 & AH Dra & SRb & O & 158 & 156 & 1.25 & 4$\pm$$^{11}_{16}$& 0.03$\pm$$^{0.07}_{0.10}$ & 0 & 0.00 \\
  &     &   &   &   &   & 2.2 & 4$\pm$$^{13}_{24}$& 0.03$\pm$$^{0.08}_{0.15}$ & $-$7 & $-$0.04 \\
  &     &   &   &   &   & 3.5 & 2$\pm$$^{14}_{22}$& 0.01$\pm$$^{0.09}_{0.14}$ & $-$11 & $-$0.07 \\
D19213546P7633345 & UX Dra$^b$ & SRa & C & 168 & 191 & 1.25 & $-$10$\pm$$^{78}_{53}$& $-$0.05$\pm$$^{0.41}_{0.28}$ & 11 & 0.06 \\
  &     &   &   &   &   & 3.5 & $-$15$\pm$$^{62}_{45}$& 0.08$\pm$$^{0.32}_{0.24}$ & 1 & 0.01 \\
D23582487P5123190 & R Cas & M & O & 430 & 437 & 1.25 & 127$\pm$$^{22}_{115}$& 0.29$\pm$$^{0.05}_{0.26}$ & $\ge$50 & $\ge$0.11 \\
  &     &   &   &   &   & 2.2 & 121$\pm$$^{22}_{110}$& 0.28$\pm$$^{0.05}_{0.25}$ & $\ge$92 & $\ge$0.21 \\
  &     &   &   &   &   & 3.5 & 121$\pm$$^{22}_{104}$& 0.28$\pm$$^{0.05}_{0.24}$ & $\ge$46 & $\ge$0.11 \\
  &     &   &   &   &   & 4.9 & 17$\pm$$^{107}_{16}$& 0.04$\pm$$^{0.24}_{0.04}$ & $\ge$39 & $\ge$0.09 \\
  &     &   &   &   &   & 12 & 17$\pm$$^{107}_{17}$& 0.04$\pm$$^{0.24}_{0.04}$ & $\ge$20 & $\ge$0.05 \\
  &     &   &   &   &   & 25 & 17$\pm$$^{107}_{27}$& 0.04$\pm$$^{0.24}_{0.06}$ & $\ge$18 & $\ge$0.04 \\
\enddata
\tablenotetext{a}{Optical light curve from \citet{percy01}}.
\tablenotetext{b}{Optical light curve from \citet{buchler04}}.
\end{deluxetable}



\end{document}